# Numerical Estimation of Frictional Torques with Rate and State Friction

Arun K. Singh[a,*], T.N. Singh[b]
[a]Department of Mechanical Engineering, Visvasvaraya National Institute of Technology, Nagpur-10, India
[b]Department of Earth Sciences, Indian Institute of Technology Bombay, Mumbai-76, India

**Abstract**

In this paper, numerical estimation of frictional torques is carried out of a rotary elastic disc on a hard and rough surface under different rotating conditions. A one dimensional spring-mass rotary system is numerically solved under the quasistatic condition with the rate and state dependent friction model. It is established that torque of frictional strength as well as torque of steady dynamic stress increases with radius and found to be maximum at the periphery of the disc. Torque corresponding to frictional strength estimated using the analytical solution matches closely with the simulation only in the case of high stiffness of the connecting spring. In steady relaxation simulation, a steadily rotating disc is suddenly stopped and relaxational angular velocity and corresponding frictional torque decreases with both steady angular velocity and stiffness of the connecting spring in the velocity strengthening regime. In velocity weakening regime, in contrast, torque of relaxation stress deceases but relaxation velocity increases. The reason for the contradiction is explained.

---

[*] Corresponding author at E-mail: aksinghb@gmail.com

# 1.Introduction

There are a variety of engineering and technological applications in which a rotary disc interacts with a hard and rough surface for example, rotary shear apparatus (Scholz, 1990), frictional brakes etc (Ibrahim, 1994). Rotary shear apparatus (RSA) which is an alternative of direct shear test is also used in the study of friction of solids such as rocks, metals, elastomers etc. (Toro et al., 2006; Gong and Osada, 2002; Niemeijer et al., 2008; Chaudhury and Chung 1997). One of the advantages of the RSA is that it can be subjected to large displacement (Scholz, 1990). Chaudhury and Chung (2007) have used the rotary set up to study friction of elastomers such as poly dimethyl siloxane (PDMS). Gong and Osada (2002) measured the frictional properties of soft hydrogels using the rotational tribometer. Amontons-Coulombs'(AC) law,which states that frictional force is proportional to normal force, is generally used to estimate total force and corresponding torque at the rotating interface of the disc. However, one of the limitations of this classical friction law is that it does not take into account the effect of velocity and area of the contact of the sliding body. Chaudhury and Chung (2007) used a power law between frictional stress and angular velocity to model the frictional behaviour of PDMS elastomers on a hard surface (silicon wafer). In the present study, frictional torque of a rotary disc is investigated with the rate and state dependent friction (RSF) model which is basically a modified form of the Amontons-Coulombs' laws (Dieterich, 1978; Ruina 1983; Marone, 1998). The modification of the AC laws is based on the experimental observations that friction of hard and rough solid surfaces (rocks, metals, hard polymers etc.) depends on time of contact as well as slip velocity of the sliding interface (Dieterich, 1978; Ruina, 1983; Marone, 1998, Persson, 2000). Further, the RSF model has found wide spread applications in the study of earthquake source mechanics (Scholz, 1990; Marone, 1998; Persson, 2000) and also in modeling of slope stability (Chau, 1994). Notwithstanding the widespread use of the RSA (Niemeijer et al., 2008; Niemeijer and Spiers, 2006; Toro et al., 2006; Mizoguchi et al., 2006&2009; Prakash and Yuna, 2008; Xu and Freitas, 1988), modeling of the RSA in the light of the RSF model is yet to be reported in the literature. Hence, modelling and simulation of a rotary disc on a frictional surface in the framework of the RSF model is the main motivation of the present article.

A brief review of the origin and further development of the RSF model was presented ( Singh and Singh, 2012). They have studied numerically frictional strength and steady relaxation of a spring-mass sliding system with the RSF model (Singh and Singh, 2012). It is established that frictional strength varies linearly as logarithm of waiting time as well as logarithm of shear velocity. These predictions are in confirmation with the experimental observations (Dieterich, 1978; Scholz, 1990; Marone 1998). It is also shown that rate of steady relaxation of a sliding block increases with stiffness and steady velocity. These predictions are also in tune with the experimental results reported by many researchers (Marone, 1997; Scholz 1990; Persson, 2000). Motivated from this study, this approach is now extended analogously to a rotary disc in the present study. It is to be noted that linear velocity, unlike the direct shear sliding case, varies with radial distance from the centre of the rotating system such as a disc. Thus, the expressions derived for the case of direct shear sliding can be replaced with $V = r\omega$ where $r$ and $\omega$ are radial distance and angular rotation of the disc respectively. However, in order to calculate total force and corresponding torque, one has to integrate all elemental stresses at the interface from centre to periphery of the disc.

Ruina (1983) proposed the RSF model in terms of an internal variable $\theta$ and instantaneous slip velocity $V$ of the sliding surface. According to this model, frictional stress $\tau$ in terms of angular velocity $\omega$ and radial distance $r$ is given as

$$\tau = \tau_* + A \ln(r\omega/V_*) + B \ln(V_*\theta/L) \qquad (1)$$

where $\tau_*$ and $V_*$ are reference frictional shear stress and reference sliding velocity respectively. Further, $A$ and $B$ are the frictional parameters and generally considered proportional to normal stress (Ranjith and Rice, 1999). As mentioned, $\theta$ represents the state of the contacting surfaces and $L$ is a critical distance over which evolution of microcontacts occurs (Ruina, 1983; Marone, 1998; Ranjith and Rice, 1999). $L$ is generally the order of the size of microcontacts (Ruina, 1983; Dieterich, 1978; Marone, 1998).

More than two empirical laws for the evolution of state variable $\theta$ have been proposed (Marone, 1998; Ranjith and Rice, 1999). However, Dieterich-Ruina aging law will be used in the present study as this law characterizes the true aging of the contacting surfaces during stationary state (Ruina, 1983; Marone, 1998) and generally expressed in terms of radial distance $r$ and angular velocity $\omega$ as following

$$\frac{d\theta}{dt} = 1 - r\omega\theta/L \qquad (2)$$

An important feature of Eq 2 is that during the stationary state $(\omega=0)$, state variable $\theta$ becomes proportional to true time of contact (Ranjith and Rice, 1999). Moreover, the above law (Eq 2) reduces to the steady-state value $\theta_{ss}$, under steady sliding, i.e, $\theta_{ss} = L/V_{ss}$ (Ranjith and Rice, 1999) where $V_{ss} = r\omega_{ss}$. Notably, the conditions for steady state sliding are $d\theta/dt = 0$ and $d\tau/dt = 0$ (Ranjith and Rice, 1999; Persson, 2000). Thus, $\theta_{ss}$ signifies average time to renew the contacts during steady sliding (Ranjith and Rice, 1999). The expression for steady dynamic stress $\tau_{ss}$ is given by (Gu et al., 1984, Ranjith and Rice, 1999)

$$\tau_{ss} = \tau_* - (B-A)\ln(r\omega_{ss}/V_*) \qquad (3)$$

From Eq 3, it is obvious that steady frictional stress $\tau_{ss}$ decreases with steady velocity $V_{ss}$ if $B > A$ and this is known as the velocity weakening process. However, in the case of $A > B$, $\tau_{ss}$ increases with sliding velocity and this is called the velocity strengthening process (Persson, 2000; Ranjith and Rice 1999; Marone, 1998).

In Fig.1, a schematic sketch of a rigid disc having radius $R$ connected with an elastic spring of rotational stiffness $K$ (per unit area of contact surface) is being rotated with a constant angular velocity $\omega_0$. The other end of the spring is attached with the disc rotating with angular velocity $\omega$. This rotating disc is, in turn, in contact with the fixed surface. Torque $M$ is needed to rotate the disc against the friction force $F$ at the rotating interface.

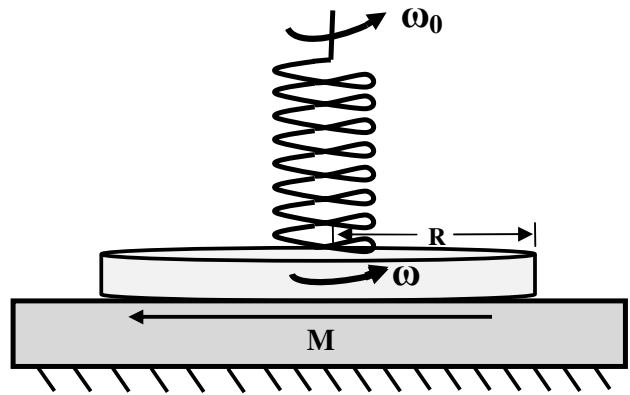

**Figure.1** A schematic sketch of the spring- rotary disc on a fixed surface.

Frictional stress $\tau$ of the spring-mass system under quasistatic (neglecting inertia of the disc) rotation is given by

$$\frac{d\tau}{dt} = Kr(\omega_0 - \omega) \tag{4}$$

In order to study waiting time $t_w$ and angular velocity (external) $\omega_0$ dependent frictional strength, a variable $X$ is defined by assuming that the state variable $\theta_w$ corresponding to waiting time $t_w$ evolves to $\theta$ as function of time $t$ as follows

$$X = \frac{(\theta_w - \theta)}{\theta_w}, \text{ or } \theta = (1 - X)\theta_w \tag{5}$$

Non-dimensional form of Eq5 is expressed using non-dimensional state variable $\hat{\theta} = \theta V_* / L$ as

$$\hat{\theta} = (1 - X)\hat{\theta}_w \tag{6}$$

Thus the non-dimensional form of Eq 2 in terms of $\hat{\theta}$ is

$$\frac{dX}{dT} = -\hat{\theta}_w^{-1} + (1 - X)\rho e^{\phi} \tag{7}$$

Eq1 is used to calculate frictional stress $\tau$ and its non-dimensional form $\psi$ is as follows

$$\psi = \psi_* + \ln(\rho) + \phi + \beta \ln\left[(1 - X)\hat{\theta}_w\right] \tag{8}$$

The final form of ordinary differential equations governing frictional dynamics of the rotary system in Fig.1 may be expressed as

$$\frac{d\phi}{dT} = k\rho(\varphi_0 - e^{\phi}) - \beta\left[\frac{1}{(1 - X)\hat{\theta}_w} - \rho e^{\phi}\right] \tag{9}$$

$$\frac{dX}{dT} = -\hat{\theta}_w^{-1} + (1 - X)\rho e^{\phi} \tag{10}$$

where non-dimensional terms are defined as $\psi = \tau/A$, $\psi_* = \tau_*/A$, $T = tV_*/L$, $\varphi_0 = \omega_0/\omega_*$, $V_* = R\omega_*$, $\phi = \ln(\omega_0/\omega_*) = \ln(\varphi_0)$, $\beta = B/A$, $k = KL/A$, $\rho = r/R$, $R$ is the radius of the circular disc. Finally, non-dimensional form of waiting time $T_w$ dependent state variable $\hat{\theta}_w$ for a rotational disc is defined as following

$$\hat{\theta}_w = \left[(\rho\varphi_0)^{-1} - kT_w/(\beta-1)\right] \tag{11}$$

The derivation of $\hat{\theta}_w$ for direct shear sliding is described in the paper (Singh and Singh, 2012). The corresponding expression for dimensionless steady rotational stress $\psi_{ss}$ is given by

$$\psi_{ss} = \psi_* - (\beta-1)\left[\ln(\rho) + \phi_{ss}\right] \tag{12}$$

It is quite easy to establish that the stability criterion of a rotary disc is the same as that of direct shear sliding system. This is because the final linearized equation governing the steady stability of a rotary disc is independent of radius of the disc.

As mentioned earlier, in order to calculate total stress or total torque at the interface, one has to sum up all stresses along the radius as stress changes with radius owing to change in linear velocity. Torque of a force is defined as the multiplication of force and distance from the centre of rotation. The expression for calculating total torque $M$ of frictional force is given by

$$M = 2\pi A R^3 \int_0^1 \psi(\rho, T)\rho^2 d\rho \tag{13}$$

Non-dimensional torque $\mu$ of frictional stress is expressed as $\mu = M/\pi A R^3$. It is to be noted that all results concerning frictional stress in the present paper will be expressed in the term of $\mu$. Further, considering the wide spread use of rotary shear apparatus, it becomes interesting to correlate friction parameter such as $\beta$ related with the RSF model in terms of measurable parameter for instance, torque $\mu$ of frictional stress in the actual rotational experiments.

## 2. Frictional Strength of a Rotational Disc

Frictional strength of an interface is defined as the minimum force required to initiate the motion of a body from rest. In frictional sliding experiments on rock surfaces, it has been observed that frictional strength of an interface depends on time of stationary contact and sliding velocity (Dieterich,1978; Marone, 1998). Eq 1 may be approximated for both waiting time and angular velocity dependent $\mu_{max}$ of frictional strength. Waiting time $T_w$ and angular velocity $\varphi_0$ dependent $\mu_{max}$ may be expressed as

$$\psi_{max} = \psi_* + \ln(\rho) + \ln(\varphi_0) + \ln\left[(\rho\varphi_0)^{-1} - \frac{kT_w}{(\beta-1)}\right] \tag{14}$$

After plugging Eq 14 into Eq 13 for estimating $\mu_{max}$ is expressed as

$$\mu_{max} = 2\psi_*/3 - 2/9 + 2\ln(\varphi_0)/3 + 2\beta \int_0^1 \ln\left[(\rho\varphi_0)^{-1} - kT_w/(\beta-1)\right]\rho^2 d\rho \tag{15}$$

An approximate solution of Eq 15 in the velocity strengthening regime $(\beta < 1)$ may be found by assuming that $kT_w/(\beta-1) \gg (\rho\varphi_0)^{-1}$ as

$$\mu_{max} = 2\psi_*/3 - 2/9 + 2\ln(\varphi_0)/3 + (2\beta/3)\left[\ln(kT_w/(\beta-1)) - 1/3\right] \tag{16}$$

It can be established using Eq 16 that slope $d\mu_{max}/\ln(\varphi_0)$ of $\mu_{max}$ vs. $\varphi_0$ is equal to 2/3 for a fixed waiting time $T_w$. Similarly, it can be also estimated using Eq16 that slope $d\mu_{max}/\ln(T_w)$ of $\mu_{max}$ vs. $T_w$ is equal to $2\beta/3$ for a fixed angular velocity $\varphi_0$. These results are in confirmation with the numerical simulation of Eq 9 and Eq10.

## 3. Steady Dynamic Stress

The expression for non-dimensional steady torque $\mu_{ss}$ is estimated with Eqs 12 and 13, the exact expression for $\mu_{ss}$ is found to be as

$$\mu_{ss} = \frac{2\psi_*}{3} - \frac{2(\beta-1)\left[\ln(\varphi_{ss}) - 1/3\right]}{3} \tag{17}$$

From Eq17, it may be concluded that slope $d\mu_{ss}/d\ln(\varphi_{ss}) = 2(\beta-1)/3$ in the velocity weakening regime $(\beta > 1)$ while value of $d\mu_{ss}/d\ln(\varphi_{ss}) = 2(1-\beta)/3$ in the velocity strengthening regime $(\beta < 1)$. Further, like direct shear sliding, in the present case too, it is obvious that $\mu_{ss}$ varies proportional to logarithm of steady angular velocity i.e., $\ln(\varphi_{ss})$ (Singh and Singh, 2012).

## 4. Steady Relaxation of a Rotating Disc

The motivation to simulate steady relaxation of a rotational disc emanates from the fact that steady relaxation is commonly observed if a dynamical system is suddenly stopped (Marone, 1998). Simulation of steady relaxation concerning direct shear sliding of spring-mass system has been carried (Singh, 2012). Thus, it becomes interesting to extend that study in the case of a rotating disc as well. Moreover, this study is of practical significance during the slide-hold-slide (SHS) experiments with a rotary shear apparatus. Many researchers (Niemeijer and Spiers, 2006; Mizoguchi et al., 2006) have carried out SHS experiments on hard surfaces with rotational shear apparatus. Derivation concerning steady relaxation velocity and corresponding interfacial stress is described in the paper (Singh and Singh, 2012). The expressions are now used in the present study for simulating steady rotational relaxation process just by replacing $V = r\omega$. Dimensionless form of relaxational angular velocity $\hat{\phi}_r$ is expressed as

$$\hat{\phi}_r = -\ln\left(e^{-\phi_{ss}} - kT_r/(\beta-1)\right) \tag{18}$$

And corresponding steady relaxation stress (dimensionless) is given as

$$\psi_r = \psi_* - (\beta-1)\left[\ln(\rho) + \hat{\phi}_r\right] \tag{19}$$

Finally, torque $\mu_r$ of relaxation stress is evaluated using Eq13 as follows

$$\mu_r = 2\psi_*/3 - 2(\beta-1)\int_0^1 \left[\ln(\rho) + \phi_r\right]\rho^2 d\rho \tag{20}$$

A parametric study of Eq 20 is carried out with respect to steady velocity $\phi_{ss}$ and stiffness $k$ of the connecting spring in both regimes namely velocity strengthening $(\beta<1)$ and velocity weakening $(\beta>1)$ to understand the steady angular relaxational process of the rotary disc.

## 5. Results and Discussion

### 5.1. Simulation of Frictional Strength

Aiming to study the effect of applied angular velocity $\varphi_0$ on torque of frictional strength in the velocity strengthening regime for $\beta = 0.2$, $\varphi_0$ is applied on the free end of the rotational spring (Fig.1) for a fixed waiting time $T_w = 1$, and $k = 1$ and $\psi_* = 1$. This simulation is repeated for different $\varphi_0 = 2, 4, 6, 8,$ and $10$ and the results are presented in Fig.2a for $\mu$ vs. $\rho$ for varying $\varphi_0$. Maximum value of $\mu$ is known as torque of frictional strength $\mu_{max}$ of the rotating interface and is found at $\rho = 1$ (Fig. 2a). This trend is expected since linear velocity $V$ will be maximum at the periphery of the disc, i.e., $r = R$. Fig.2 b, on the other hand, presents $\mu_{max}$ vs. $\varphi_0$ for varying stiffness $k = 0.5, 1.0 \,\&\, 10.0$ of the connecting spring. It may be seen that $\mu_{max}$ varies linearly with logarithmic of $\varphi_0$ for a fixed waiting time $T_w = 1$. Slope $d\mu_{max}/d\ln(\varphi_0)$ of the best fit lines in Fig.2b is found to be equal to $0.620, 0.659 \,\&\, 0.666$ for stiffness $k = 0.5, 1 \,\&\, 10$ respectively. It is concluded from Fig.2b that slope $d\mu_{max}/d\ln(\varphi_0)$ increases with stiffness $k$ and approaches to nearly equal to $0.666$ in the case of high stiffness of the connecting spring.

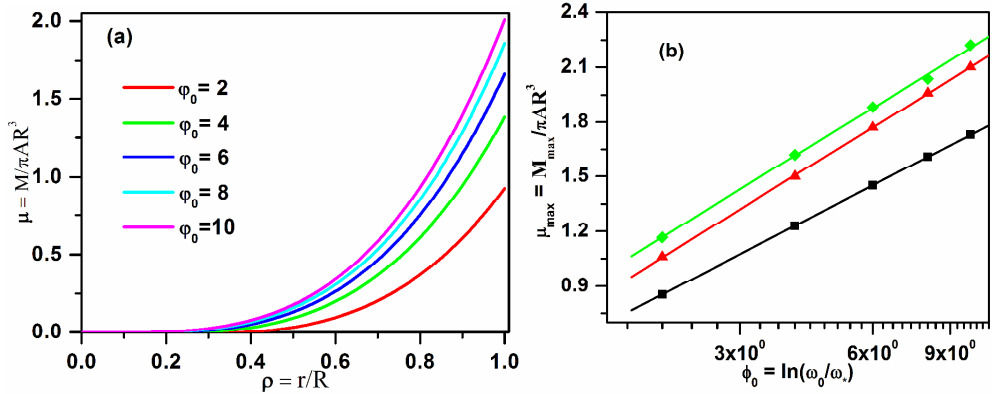

**Fig.2.** presents (a) $\mu$ vs. $\rho$ and (b) $\mu_{max}$ vs. $\varphi_0$ for fiction parameters $\beta = 0.2$, $k = 1$ and $\psi_* = 1$ for varying stiffness $k = 0.1$(black), $k = 1$ (red), and $k = 10$(green).

Waiting time $T_w$ dependent $\mu_{max}$ of the rotating disc is also studied in the velocity strengthening regime $(\beta < 1)$. In this simulation, $T_w$ is changed at a fixed applied angular velocity $\varphi_0 = 2$ for different stiffness $k = 0.5, 1.0, \,\&\, 10.0$ of the connecting spring. In $T_w$ dependent $\mu_{max}$ simulation results in Fig.3, the trend of $\mu_{max}$ vs. $\rho$ is found to be the same trend as shown in Fig.2a. It is also seen in Fig.3b that $\mu_{max}$ varies linearly with $\ln(T_w)$. Moreover, friction parameter $\beta$ estimated from the slope $d\mu_{max}/d\ln(T_w) = 2\beta/3$ of the best fit lines in Fig.3 is estimated to be equal to $0.158, 0.188 \,\&\, 0.195$ for stiffness $k = 0.5, 1 \,\&\, 10$ respectively. These results also confirm that numerical value of $\beta$ obtained using the simulations matches with the assigned value of $\beta$ i.e., $\beta = 0.2$ only in the case of high stiffness of the spring.

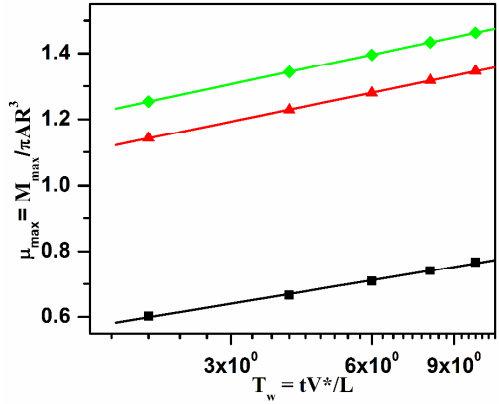

**Fig.3.** presents $\mu_{max}$ vs. $T_w$ in the velocity strengthening regime for friction parameters $\beta = 0.2$, $\psi_* = 1$ and $T_w = 1$ for varying stiffness $k = 0.1$(black), $k = 1$ (red), and $k = 10$(green).

Validity of analytical expression in Eq 16 for $\mu_{max}$ is investigated with the simulated results. The results are illustrated in Fig.4 ($\mu_{max}$ vs. $\phi_0$) for a fixed waiting time $T_w = 1$. Fig. 4 also contains the results for different stiffness $k = 0.1, 5,$ and $10$. At the same time in Fig.4, each colour showing two straight lines correspond to the solution obtained through analytical (upper) and simulation (lower) approaches. More interestingly, it may be noted from the plots in Fig.4 that there is a considerable difference between analytical expression in Eq16 and simulation for instance, $k = 0.5$

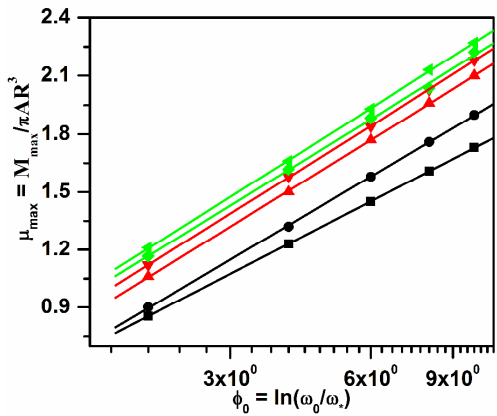

**Fig.4** presents $\mu_{max}$ vs. $\phi_0$ for friction parameters $\beta = 0.2$, $\psi_* = 1$ and $T_w = 1$ for varying stiffness $k = 0.1$ (black: simulation(lower), analytical (upper) ), $k = 5$ (red: simulation (lower), analytical(upper)), and $k = 10$(green: simulation(lower), analytical (upper)).

However, difference between two approaches (analytical and simulation) reduces as stiffness of the connecting spring increases to $k = 1$ and 10. Hence, on the basis of this result, it is claimed that the analytical solution for $\mu_{max}$ (Eq16) is valid only in case of stiff spring say $k > 10$.

5.2. Simulation of Steady Dynamic Frictional Stress

Numerical simulation of torque of steady stress $\mu_{ss}$ is also carried out with respect to angular velocity $\phi_{ss}$. One of the objectives for steady dynamic simulation is that dynamic friction as a function of angular velocity is widely used to determine the friction parameter $\beta$ in friction experiments. The results are plotted in Fig. 5a for velocity strengthening regime. Plot in Fig.5b shows that variation of $\mu_{ss}$ varies linearly with $\ln(\varphi_{ss})$. Slope $d\mu_{ss}/d\phi_{ss}$ of the best fit line in Fig.5b is found to be equal to 0.533. Moreover, $\beta$ obtained from the slope of the best fit line is equal to 0.2. This result confirms the validity of the value of the slope $d\mu_{ss}/d\phi_{ss} = 2(1-\beta)/3$ of $\mu_{ss}$ vs. $\ln(\varphi_{ss})$ in the velocity strengthening regime.

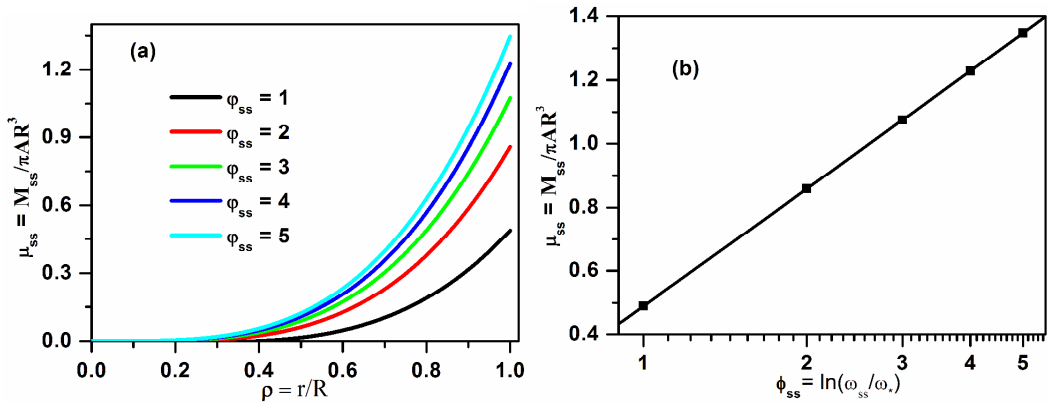

**Fig.5.** shows (a) $\mu_{ss}$ vs. $\rho$ and (b) $\mu_{ss}$ vs. $\phi_{ss}$ for friction parameters $\beta = 0.2$, $k = 1$ and $\psi_* = 1$. Slope of the best fit line is found to be equal to 0.533.

In the velocity weakening regime $(\beta > 1)$ too, similar simulation is carried out for $\beta = 1.2$ for varying $\phi_{ss}$. The results are presented in Fig. 6a which shows that $\mu_{ss}$ decreases with increase in angular velocity $\phi_{ss}$. Further, Fig.6b presents $d\mu_{ss}/d\phi_{ss}$ vs. $\phi_{ss}$ and its value found to be equal to $1.2$. This results in confirmation with the slope value $d\mu_{ss}/d\phi_{ss} = -2(\beta-1)/3$ which is obtained from the analytical solution in Eq17.

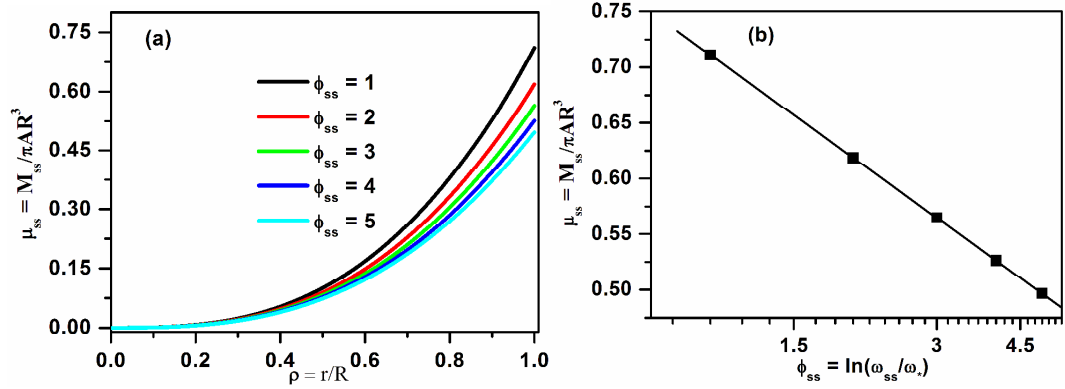

**Fig.6.** shows (a) $\mu_{ss}$ vs. $\rho$ and (b) $\mu_{ss}$ vs. $\phi_{ss}$ for friction parameters $\beta = 1.2$, $k = 1$ and $\psi_* = 1$. Slope of the best fit line is -0.133.

5.3. Simulation of the Steady Relaxation in Velocity Strengthening Regime $(\beta < 1)$

Similar to the steady relaxation of shear sliding (Singh and Singh, 2012), simulation of steady relaxation of the rotary disc is also carried out. Relaxation time $T_r$ dependent relaxation of interfacial velocity $\phi_r$ and corresponding torque $\mu_r$ is also studied as function of $\phi_{ss}$ and stiffness $k$ of the spring. Eqs18 & 19 are plotted for $\beta < 1$. Figs.7a present the results concerning relaxation velocity $\phi_r$ vs. $T_r$ and Fig.7b contains the plots between $\mu_r$ vs. $T_r$ for different $\phi_{ss} = 0.01, 0.1 \& 1.0$. It may be seen in the plots that both $\phi_r$ and corresponding $\mu_r$ decrease with relaxation time $T_r$ for all $\phi_{ss}$. This is also observed in experiments when the steadily sliding mass is suddenly brought to a halt (Marone, 1997). It may also be inferred from the plots in Figs.7a&b that initially the rate of relaxation velocity $\phi_r$ and corresponding torque $\mu_r$ of relaxation stress both increase with relaxation time but the rate becomes the same later irrespective of $\phi_{ss}$.

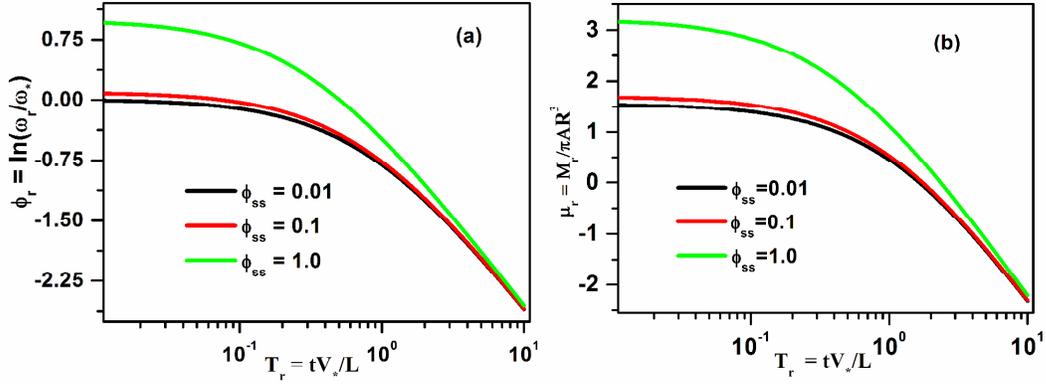

**Fig.7.** Effect of steady velocity $\phi_{ss}$ on relaxation process in the velocity strengthening regime $(\beta = 1.2)$: (a) relaxation velocity $\phi_r$ vs. $T_r$; (b) relaxation stress $\mu_r$ vs. $T_r$ for varying $\phi_{ss} = 0.01$, $0.1$ and $1.0$.

Fig.8a shows relaxation velocity $\phi_r$ and $\mu_r$ (Fig.8b) at the sliding interface as function of relaxation time $T_r$ for varying stiffness $k = 0.01, 0.1 \& 1.0$. It is clearly visible that $\phi_r$ decreases with stiffness $k$ of the connecting spring. Similarly in Fig.7b, $\mu_r$ decreases with relaxation time $T_r$ as well. These observations are similar to the relaxation results in direct shear sliding (Singh and Singh, 2012).

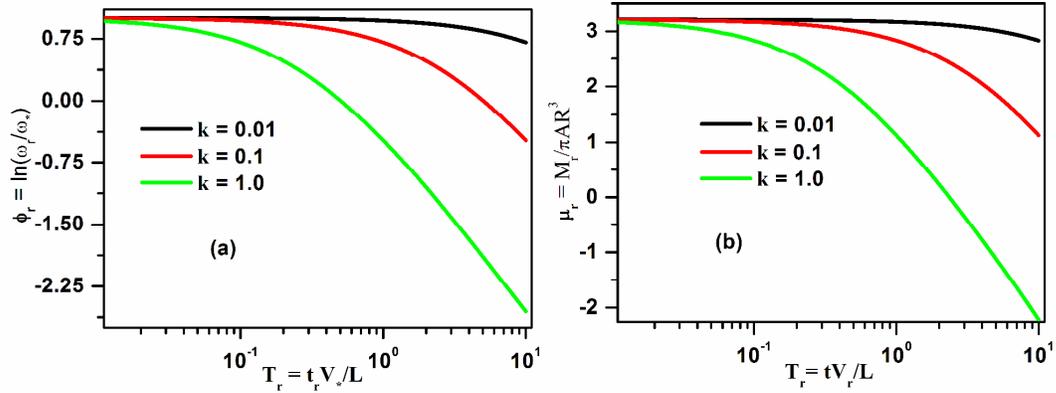

**Fig.8.** Effect of spring stiffness $k$ on relaxation process in the velocity strengthening regime $(\beta = 0.2)$: (a) relaxation velocity $\phi_r$ vs. $T_r$; (b) relaxation stress $\mu_r$ vs. $T_r$ for varying $k = 0.01$, $0.1$ and $1.0$.

### 5.4. Simulation of the Steady Relaxation in Velocity Weakening Regime $(\beta > 1)$

Numerical simulations are also carried out to study the steady relaxation of the rotary disc in the velocity weakening regime $(\beta = 1.2)$. The results are presented in Figs.9a &9b demonstrate the effect of $\phi_{ss}$ on $\phi_r$ and $\mu_r$ respectively. It is interesting that $\phi_r$ increases but corresponding $\mu_r$ decreases with time $T_r$ for all $\phi_{ss} = 0.01, 0.05 \& 0.1$. This observation is, in contradiction, with the relaxation process in the velocity strengthening regime (Figs.7a &8a). The reason for this observation may be attributed to the velocity weakening effect in which frictional stress decreases with increase in velocity (Persson, 2000). Persson has attributed this effect to the creep of microcontacts in the perpendicular direction of sliding motion which dominates over the simultaneous creep process in the parallel direction of motion, thereby, leads to velocity weakening effect(Persson,2000).

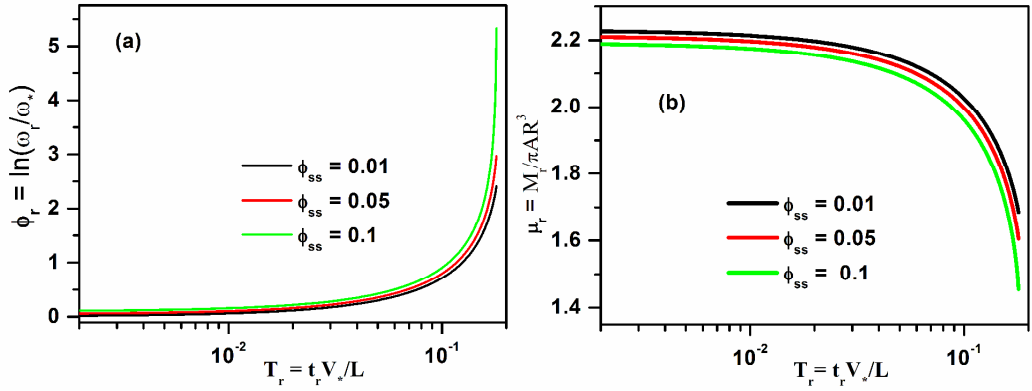

**Fig.9.** Effect of steady velocity $\phi_{ss}$ on relaxation process in velocity weakening regime $(\beta = 1.2)$: (a) relaxation velocity $\phi_r$ vs. $T_r$; (b) relaxation stress $\mu_r$ vs. $T_r$ for varying $\phi_{ss}$ = 0.01, 0.05 and 0.1.

The effect of stiffness $k$ of the connecting spring is also investigated on the relaxation process for $\beta = 1.2$. The results are plotted in Figs.10 a and b. It is inferred from the plots that increasing $k$ results in increase of $\phi_r$ (Fig. 10 a). This observation is also correct for corresponding relaxation torque $\mu_r$ (Fig. 10 b).

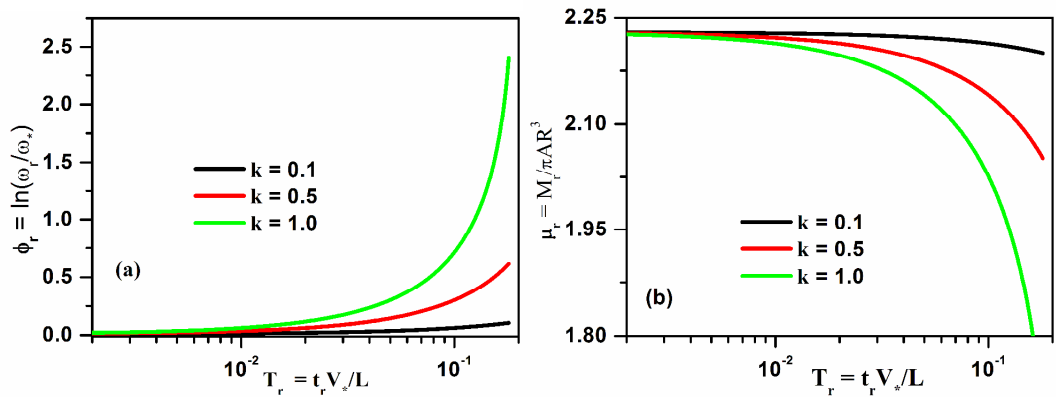

**Fig.10.** Effect of spring stiffness $k$ on relaxation process in the velocity weakening regime $(\beta = 1.2)$: (a) relaxation velocity $\phi_r$ vs. $T_r$; (b) relaxation stress $\mu_r$ vs. $T_r$ for varying stiffness $k$ = 0.1, 0.5 and 1.0.

This observation is similar to rate of relaxation process in velocity strengthening regime as mentioned in Figs 7&8. In this case also, it may be seen in the plot in Fig. 10a that $\phi_r$ increases but corresponding $\mu_r$ decreases with $T_r$. The reason is again attributed to the velocity weakening effect as explained earlier that friction decreases with increase in sliding velocity. It would be also interesting to validate the above simulated results using rotational shear apparatus in slide-hold-slide experiments.

## 6. Conclusions

In this article, numerical estimation of frictional torque of a rotary disc is carried out on a hard and rough surface under the quasistatic conditions. It is established that the torque of frictional strength as well as torque of steady dynamic stress increase with radius and found to be maximum at the periphery of the disc. Torque of frictional strength varies proportionally to logarithm of shear velocity as well as logarithm of waiting time. More significantly, the approximate expression developed for frictional strength matches with simulated one only in case of high stiffness of the connecting spring. Torque of steady dynamic stress also varies as logarithm of shear angular velocity in both velocity weakening and in velocity strengthening regimes. In steady relaxation, it is found that rate of relaxation of the frictional torque increases with both steady velocity and stiffness of the connecting spring in both regimes of frictional sliding. Moreover, in the velocity weakening regime, relaxational angular velocity increases but corresponding frictional torque decreases with time. This observation is contradictory to the velocity strengthening regime in which both relaxational angular velocity and corresponding relaxation torque decreases with time.